\documentclass[preprint]{emulateapj}

\begin{document} 
\title{\bf The Case of the Off-Center, Levitating Bar in the
Large Magellanic Cloud}

\author{Dennis Zaritsky}
\affil{Steward Observatory, Univ. of Arizona, Tucson, AZ, 85721, email:
dzaritsky@as.arizona.edu} 

\begin{abstract}
I explore the hypothesis that many of the unusual aspects of the apparent stellar bar of
the Large Magellanic Cloud are the result of viewing a triaxial
stellar bulge that is embedded in a highly obscuring thin
disk. Specifically, this hypothesis explains
the observed off-center position of the bar within the disk, the differing apparent 
distances of the bar and disk, the near alignment of the bar's major axis position
angle and the disk's line-of-nodes, 
and the asymmetric appearance of the bar itself. Indirectly, it may also play a role in 
explaining the microlensing rate toward the LMC and the recently observed
large velocity dispersion of RR Lyrae stars.
\end{abstract} 


\section{Introduction}

The off-center bar of the Large Magellanic Cloud (LMC) has helped define a class
of irregular galaxies \citep{dev} and become a prime motivation for the study of 
off-center, spiral wave
driving models \citep{ca,gardiner}. Despite the obvious ``barred" appearance of
the LMC (Figure \ref{images}a), 
there are surprisingly few direct constraints on the three dimensional structure of
the apparent feature \citep[see][]{ze}. Is the bar in the LMC real or an illusion?

Several features of the LMC bar are difficult 
to reconcile with the simplest version of barred disk galaxies:

\noindent
1) The bar is off-center relative to the underlying disk.  
At large radius ($R > 5$ kpc), where there is no contamination of the disk isophotes
by other components and where the deprojected position angle, ellipticity, and 
centroid remain approximately constant, the disk center is measured to lie approximately at
$\alpha=5^h29^m {\rm and\ } -69^\circ 30^\prime $ \citep[see Fig 4 of][]{marel}.
At small to intermediate radii (0.5 kpc $< R <$ 2.5 kpc), where the bar dominates
the isophotes and the deprojected position angle and centroid remain approximately 
constant, the centroid is shifted from that of the outer disk by $\sim$ 0.5 and 0.3 kpc
in $x$ and $y$, respectively \citep{marel}.
This type of misalignment has been 
defined to be a common characteristic of barred Magellanic irregulars \citep{dev} and dynamically
important in defining the weak spiral structure of these systems, but
its cause has not yet been conclusively identified \citep{gardiner}.

\noindent
2) The measured distance to the bar is different than that to the disk.  \cite{nik04} conclude
that the bar is ``elevated" above
the disk plane, located closer to us by $\sim 0.5$ kpc. 

\noindent
3) The bar's major axis is nearly parallel to the disk's line-of-nodes ($125^\circ\pm 5^\circ$
and $122.5^\circ \pm 8.3^\circ$, respectively; \citet{marel}).
The foreshortening of any bar whose major axis position angle
is not oriented along the line-of-nodes 
will make the alignment appear closer than it truly is, but most bars
in disk galaxies
are sufficiently elongated that they are typically not  aligned in projection.

\noindent
4) The bar is asymmetric. The stellar density is ellipsoidal on the
southwestern side and
flat along the northeastern edge. The flat edge lies nearly 
parallel to the disk's line-of-nodes
(Figure \ref{images}b). 

The hypothesis I explore
is that the LMC's barred appearance and the features listed above are the result of viewing
a bulge population that is obscured by an optically thick inner disk.

\section{A Simple Model}

\subsection{The Observational Situation}

The data come from the Magellanic Clouds
Photometric Survey, which is a $U, B, V, $ and $I$ survey
of the central $8.5^\circ \times 7.5^\circ$ of the Large Magellanic Cloud
and $4.5^\circ \times 4^\circ$ of the Small Magellanic Cloud. Both catalogs
are  published \citep{zar02,zar04}.

\begin{figure*}
\plotone{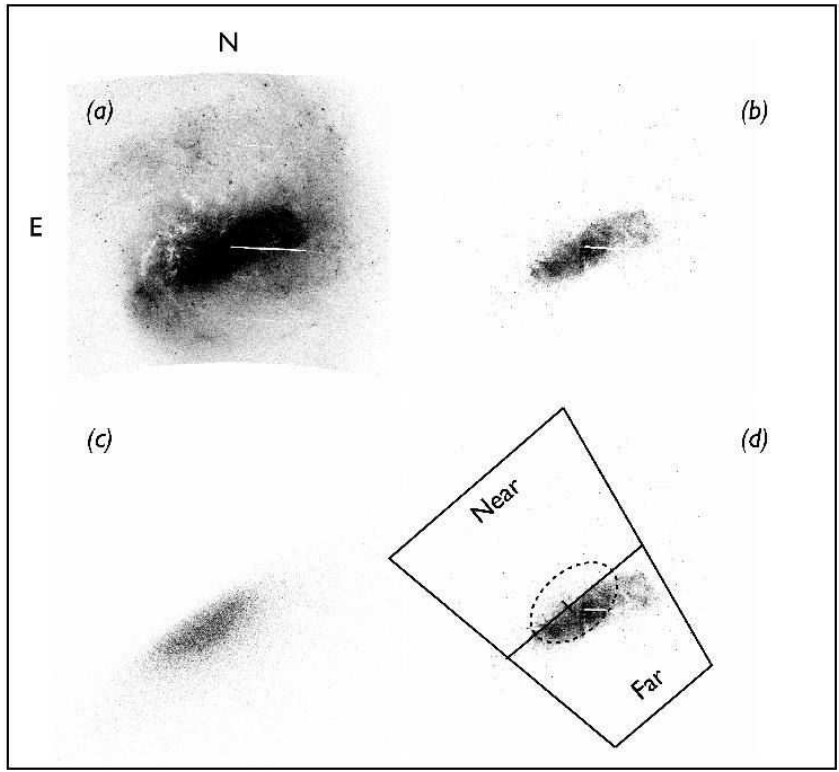}
\figcaption{The observed and modeled stellar density distribution in the LMC. Panel (a)
is a low contrast map of the stellar density for stars with $16 < V < 20$. 
Panel (b) is a high contrast map of the same stellar density distribution. There is a
central population that is sharply truncated toward the northeast, rounded toward
the southwest, and roughly aligned with the disk's line-of-nodes. Panel (c) shows the
distribution of bulge particles in the model described in the text. Panel (d) is the 
same high contrast map shown in (b) with an overlay indicating the geometry of the
disk and bulge system. The line-of-nodes (from \cite{mc}) is represented by
the straight line that bisects the rectangle, and the position of LMC center as
measured from the outer isophotes is marked with a tick mark perpendicular to the 
line-of-nodes that is slightly off-center relative to the ellipsoid that delineates the
bulge appearance in the absence of the optically thick disk.
\label{images}}
\end{figure*}

I plot in Figure \ref{images} the stellar density on the sky of stars with
$16 < V < 20$. The high-contrast panels (Figures \ref{images}b,d), show how the 
distribution of the central population appears to be sharply truncated toward
the northeast and rounded toward the southwest. The line-of-nodes and
center derived by \cite{marel} are plotted for reference in Figure \ref{images}d. 
Some rectilinear irregularities in the density distribution that are present in Figure \ref{images} 
arise from different completeness
limits among drift scan images in this region of high stellar density. Although
these irregularities do
not affect the qualitative conclusions presented here, they do preclude
detailed model fitting to these data prior to obtaining the results from 
extensive artificial star tests. I do, however, provide $\chi^2_\nu$ values for 
the various models presented for intercomparison, assuming only Poisson
noise in the star counts. These 
values are calculated within a rectangular
region than incorporates the visible bar in Figure \ref{images}b, and the model
is normalized to have the same total number of stars within this region. Because of the systematics
described above, and the lack of a modeled disk component, $\chi^2_\nu$ is 
expected to be $\gg 1$ even for the correct model. 

\subsection{Exploring the Hypothesis}

I begin with a model defined by the disk geometry
measured by \cite{mc}. The choice of geometry, in particular the position angle of
the line-of-nodes (122$^\circ$ in that study), 
will be explored below. Within the adopted geometry, the bulge 
has two of its principal
axes in the disk plane (and hence the third perpendicular to the disk plane), but
has an otherwise unconstrained position angle and axial ratio. Its center is assumed to 
be coincident with that of the disk, because the focus here is to determine whether
the observations can be explained without invoking an offset component.
I assume that all of the obscuring dust is in the disk plane and has a much smaller
vertical scaleheight than any
dynamically hot population \citep[see][for a confirmation of this assumption
for the LMC as a whole]{zar04}. The volume density of bulge stars is assumed to 
be constant for $0.001 < r \le  0.3$, where $r \equiv \sqrt{(x/a)^2 + (y/b)^2 + (z/c)^2}$, 
to follow an $r^{-3}$ density profile for $0.3 < r < 4$, 
and to otherwise be zero.
The scales, $a, b,$ and
$c$, are free parameters. The constant density core is set to avoid a divergence
at small radii and to mimic the observed shallow density gradient in the inner kpc. Because of
crowding in the MCPS images, the true profile may be significantly steeper
than that suggested by Figure \ref{images} and hence, the flat central density of the models
may not reflect reality even though it is ultimately successful in reproducing the observations.
I draw stars from a power law luminosity function, 
$N(m) \propto m^{2}$, place them in projection
according to the viewing geometry of the LMC, apply extinction (extinction
prescription is discussed below), and accept only those stars that satisfy the
observational criteria. Distance differences are accounted for in determining the apparent magnitude and no disk component is included in the generated stellar density maps.
In summary, the parameters of the model
that vary are the three scaling parameters, $a, b, $ and $c$, the core radius in scale-free
units (although
it was fixed for all models once a satisfactory value was found), and the bulge's position angle.

\begin{figure}
\plotone{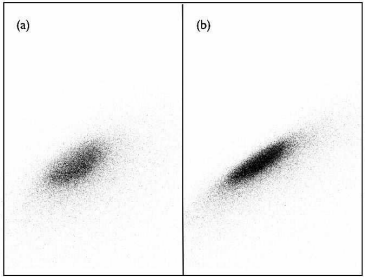}
\figcaption{The effect of changing the y and z bulge scales. Panel (a) shows a model
with $a = 2.5, b = 1.0$, and $c = 2.5$. Panel (b) shows a model with $a =2.5, b = 0.5$, and
$z = 1.5$. For comparison, the model in Figure \ref{images} has $a =2.5, b = 0.5$ and $c = 2.5$.
All of these models adopt the \cite{mc} geometry. The image is cropped to only show the inner region of the LMC.
\label{parameters}}
\end{figure}

After exploring various parameter combinations, I reached a few principal conclusions.
First, the extinction in the midplane, at least in this central portion of the LMC disk, 
must be significantly larger than estimated
by \cite{zar04}  to sufficiently obscure the rear portion of the bulge. To increase the
extinction in the modeling, I define a minimum visual 
extinction value, $A_{min}$. 
At any location for which the extinction value from
the \cite{zar04} 
extinction map is less than this minimum value, the extinction is set to $A_{min}$.
An extreme value of $A_{min} = 10$ was used in
Figure \ref{images}c to mimic complete extinction, 
but a strong extinction edge is seen toward the northeast side
of the bulge for $A_{min} > 2$. 
These large extinction values do not contradict the \cite{zar04} results because 
such highly obscured stars would not be in their sample.
Second, the extension of the bulge must be relatively short along the y-axis (the
disk plane, short axis of the bulge) because otherwise the sharp northeastern edge
created by the disk bisection of the bulge becomes increasingly blurred 
(see Figure \ref{parameters}a for a model with double the y-scale of that shown in
Figure \ref{images}c). Third, the extension of the bulge along the z-axis (out of the disk plane)
must be relatively large to produce the noticeable rounding of the southwestern edge
(Figure \ref{parameters}b). 

 The parameters for the model shown in Figure \ref{images}c
are $a = 2.5$ kpc, $b = 0.5$ kpc, and $c = 2.5$ kpc, which makes this bulge an oblate
spheroid that protrudes from the disk plane (this model has $\chi^2_{\nu} = 9.9$, while the
same geometric model without any midplane extinction has $\chi^2_{\nu} =  46.9$). 
The physical scale of the constant
density central region along each axis is given by the core boundary,
defined previously to be $r = 0.3$, times either $a,  b$, or $c$. 
The extreme nature of this bulge can be mitigated somewhat, at the expense of a
moderate worsening of the fit. For example, a bulge with
$a = 2.5$, $b = 1.0 $ and $c = 2.5$ still reproduces the principal qualitative structure
but does not produce as sharp an edge toward the northeast
(Figure \ref{parameters}a). This model has a slightly lower $\chi^2_\nu$,  9.2, than the
previous model presumably
because it is more extended and so partially accounts for the missing disk stars in the model.
The extent along the z axis cannot be decreased
significantly without losing the rounded appearance of the bar (see Figure \ref{parameters}b
for a model with $a = 2.5, b=0.5,$ and $c = 1.5$; $\chi^2_\nu = 76.8$). 
Any detailed exploration of the parameter space would require a disk model
in which the bulge is embedded, and higher resolution imaging to produced stellar 
density distributions that are less severely affected by crowding.

\begin{figure*}
\plotone{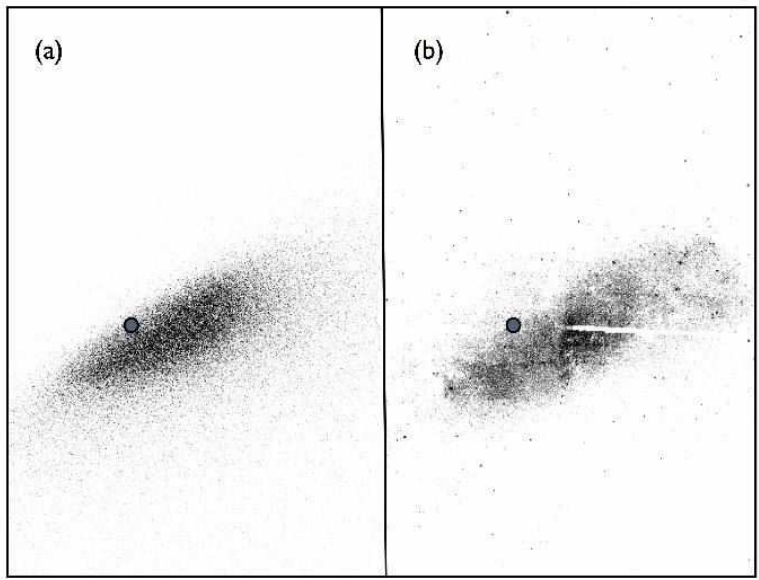}
\figcaption{Comparison of model (a) and data (b) if the disk line-of-nodes is at 150$^\circ$. 
Filled circles denote the center of the LMC as determined from the outer disk isophotes
\citep{marel}. The image is cropped to only show the inner region of the LMC.
\label{offset}}
\end{figure*}

Finally, I explore the effect of changing the adopted position angle of the disk's
line-of-nodes. The initial angle in the models was that 
measured by \cite{mc}, but another recent
measurement produced an angle that is $\sim$ 30$^\circ$ larger \citep{nik04}.
Surprisingly, this difference does not affect the model's 
ability to recreate the bar appearance. By 
changing the bulge position angle from being 10$^\circ$ off the line-of-nodes
to 45$^\circ$,  I reproduce the observed appearance (Figure \ref{offset}; $\chi^2_\nu = 10.1,
a= 3.0, b = 0.5,$ and  $c = 3.0$).
This new  geometry helps explain one feature that was not well accounted
for in the original models. The circles in Figure \ref{offset} represent the center of 
the LMC as determined from the outer disk isophotes. 
The observed midpoint of the ``bar" along the line-of-nodes is displaced by
15 arcmin toward the northwest, while the stellar density weighted centroid of
the model presented here is displaced by 23 arcmin in the same northwesterly direction.
In the simulations, at least, this offset from the center is due to the longer
line of sight through the bulge to the disk on one side of the LMC center than on the other.
The uncertainties in determining a precise LMC center
and in the stellar density distribution of the MCPS in this highly crowded region
preclude one from inverting the model fits to reach a conclusion on the correct position 
angle of the line-of-nodes. 
Nevertheless, is is possible to reproduce the qualitative nature of the 
observations over the range of observed values for the position angle of the disk's 
line-of-nodes. The relative ease with which the general model was able to fit
the data despite a large change in the line-of-nodes demonstrates that its ability
to produce populations that mimic the observations is not predicated on a special
coincidence of paramaters.

\section{Discussion and Summary}

Although 
this modeling exercise demonstrates that the LMC {\sl might} have a stellar bulge that when viewed
in projection across an optically thick disk produces a feature that appears to be 
an off-center, levitating bar,  it does not {\sl prove} that this is indeed the case. 
The principal differences
between a bar and a bulge in the LMC lie in the line-of-sight extent and the vertical (out of the disk
plane) velocity dispersion. There are tantalizing observations that suggest that
these might both be large, and hence support the bulge hypothesis. First, aside from the
different distance measurement to the disk and bar \citep{nik04}, a protruding stellar population would help
reproduce the larger-than-expected microlensing rates toward the LMC
\citep{ze}.  Although \cite{ze} speculated that the lensing arose from an unrelaxed stellar population that in projection appeared to be a bar, 
the hypothesized bulge will produce an analogous lensing enhancement 
and the strong disk-plane extinction would 
address the spatial asymmetry of the discovered microlensing events \citep{mancini}. 
Whether this hypothesis
survives further, more quantitative, scrutiny depends on the mass and radial profile of the bulge, 
both of which are unknown, but evaluating models similar to those carried out 
by \cite{mancini} for this newly proposed geometry is the next step. 
\cite{alves}
has fit King models to the distribution of RR Lyrae stars,
which presumably trace an old population, and 
finds a core radius of $1.42 \pm 0.12$ kpc (a rough match to the core
radii of 0.7 to 1.0 kpc imposed in these models),
and a steep decline (exponential scale length of $1.47 \pm 0.08$ kpc). This measurement might
be providing a description of the radial distribution of the bulge population.
Second, there is evidence for 
a kinematically hot stellar component in the LMC \citep{minniti, borrisova}. Those
authors suggest that this component is
associated with the stellar halo, but because they observe the high velocity
stars at small projected radii (for example, $R< 1.5^\circ$ for the \cite{borrisova} study),
rather than at radii where only LMC ``halo"  would exist, these stars may instead
be members of a dynamically hot bulge component.
The large velocity dispersion
\citep[$\sim 53$ km/sec;][]{borrisova}
implies a significantly larger vertical scaleheight than that of the disk for this population
\citep[the kinematically hottest disk component
has a velocity dispersion of $\sim$ 20 km/sec;][]{meath,marel02}.

Now I revisit each of the ``anomalous" observations discussed in \S1 in the context
of the bulge plus obscuring disk hypothesis:

\noindent
1) The apparent bar is off-center because we only see half the bulge, and that half at an
odd angle. The general geometrical features of the apparent bar are reproduced
well with either the \cite{mc} or  \cite{nik04} disk geometry (Figures \ref{images} and
\ref{offset}). The latter does slightly
better at fully reproducing the apparent misalignment of disk and bar centroids.

\noindent
2) The bar and disk appear to be at different distances because the bar stellar sample
is biased in favor of stars in front of the disk.
The apparent average displacements in distance of the entire bar population relative to
the disk center for
the models presented in Figures \ref{images}c
and \ref{offset}a are 1 and 0.6 kpc, respectively.
These are both larger than the measurement presented by \cite{nik04} of $\sim 0.5$ kpc,
but there are various simple ways in which the models could reproduce the published value. 
For example,
if the mid-plane extinction is reduced slightly then some stars behind the disk will
enter the sample.

\noindent
3) The bar major axis is nearly aligned with the disk line-of-nodes because the
obscuring sheet cuts through the bulge along the line-of-nodes. As such,
the alignment of the projected major axis 
with the line-of-nodes is less problematic in this obscured bulge model than in an
unobscured bar model.
For the \cite{mc} geometry, the bulge major axis in our model is within $\sim 10^\circ$ of
the line-of-nodes, but for the \cite{nik04} geometry the difference in
position angles is $\sim 45^\circ$. Within the range of observationally allowed line-of-node
angles, the model does not require precise alignment between the line-of-nodes
and bulge position angle to reproduce the observations.

\noindent
4) The apparent bar asymmetry arises because the northeastern side is
artificially truncated by the obscuring disk
(Figure \ref{images}).

Despite these successes, 
the bulge hypothesis is not without challenges. First, the shape of the derived bulge,
with a long or intermediate axis out of the plane, is unusual. 
This geometry is driven by having to reproduce the
extent of the bar in the southwestern direction. If the disk plane is actually more highly inclined to
the line of sight than in the models, then 
the bulge would not need to extend as far out of the disk to project 
as far on the sky. Because the inclination is well determined for the outer disk, such an adjustment
in inclination would require the LMC disk to be warped. 
Second, the bar feature does not appear grossly different in infrared stellar density images
\citep{marel}. For extinction to strongly affect the 2$\mu$m isopohotes would require an optical extinction
$>$ 10 \citep[$A_K/A_V \sim 0.1$;][]{card}. Such a large, rather uniform extinction appears
to be unlikely. Again, a more highly inclined disk would help the situation because the
path length through the disk would be longer and the extinction would be larger. 
Lastly, existing reddening measurements \citep{nik04,zar04} do not find a significant
tail of moderately extincted ($1 < A_V < 2$) stars. Unfortunately, the \cite{nik04}
Cepheid sample is sparse in the region of interest immediately northeast of the bar
(consisting of only a few tens of stars), the \cite{zar04} photometry is strongly affected by 
crowding in the inner LMC, and
the extinction may be sufficient to eliminate the highly extincted stars from these
samples in the first place.

How can we make further progress in determining whether the obscured bulge hypothesis
is correct? A complete kinematic survey of the LMC from which one
could determine whether the dynamically hot component has  a similar spatial distribution as the
projected bulge and whether, or not,  there are kinematic signatures within the disk of a 
bar potential could provide a definitive test.  \cite{zhao} present an extensive kinematic
survey, but for this purpose their velocity resolution is too coarse  (15 km sec$^{-1}$) and the
bar region is undersampled. Infrared surveys should eventually penetrate
through the dust and provide evidence for or against a hidden bulge population.
And so, on the basis of the available data, I can only suggest that the conundrums posed
by the LMC bar may be 
resolved by the presence of a triaxial bulge and an optically thick inner disk, and look
forward to learning whether this is indeed the case.

\begin{acknowledgements}
DZ thanks Paul Hodge and Roeland van der Marel for comments on a preliminary
draft and acknowledges financial support for his Magellanic
Cloud program from 
NSF grants (AST-9619576 and AST-0307482) and fellowships from
the David and Lucile Packard Foundation and the Alfred P. Sloan
Foundation.
\end{acknowledgements}

\end{document}